\documentstyle[12pt]{article}

\begin{document}
\title{Exact energy eigenvalues of the generalized Dirac-Coulomb
equation via a modified similarity transformation}
\author{ Omar Mustafa \\
 Department of Physics, Eastern Mediterranean University\\
 G. Magusa, North Cyprus, Mersin 10 - Turkey\\
 Thabit Barakat \\
 Department of Civil Engineering, Near East University\\
 Lefko{\c s}a, North Cyprus, Mersin 10 - Turkey\\
\date{}\\}
\maketitle
\begin{abstract}
{\small With the aid of a modified similarity transformation we have
obtained exact energy eigenvalues of the generalized Dirac - Coulomb
equation. This equation consists of the time component of the Lorentz
4-vector potential $V_{v}(r)=-A_{1}/r$, and a Lorentz scalar potential
$V_{s}(r)=-A_{2}/r$. 
The transformed radial equations are so simple so that their
solutions are inferred from the conventional solutions of the
Schr$\ddot{o}$dinger - Coulomb equation.}
\end{abstract}
\newpage
\renewcommand{\thesection}{\Roman{section}}

\section{Introduction}

Recently, Su [1] has used a simple similarity transformation to bring
the radial wave equation of Dirac - Coulomb problem into a form nearly
identical to those of the Schr$\ddot{o}$dinger and Klein - Gordon equations.
With the aid of the confluent hypergeometric functions, he was able to
come out with the exact Sommerfeld - Dirac discrete spectrum. Wong [2]
has followed the same procedure to obtain the exact solution of the
N - dimensional Dirac - Coulomb equation. The similarity transformation
S they have used was obtained by Biedenharn [3] and Wong and Yeh [4].

However, a scalar interaction is of great importance in the context of
the relativistic quark model. It is employed for describing the
magnetic moment [5] and avoiding the Klein paradox risen from the
quarkonium confining potentials [6,7]. Therefore, the search for exact
solutions to problems concerned with the scalar interaction is of
special significance. Tutik [8] has obtained an exact solution to the
bound states problem for the N - dimensional generalized Dirac -
Coulomb equation, whose potential contains both the Lorentz 4 - vector
and Lorentz scalar terms of Coulomb form. Tutik has considered only
the positive energy solution, i.e. the particle case.

In this paper we modify the similarity transformation used by Su [1]
and Wong [2] to simplify the generalized Dirac - Coulomb equation and
bring it into a form almost identical to those of the Schr$\ddot{o}$dinger
and Klein - Gordon equations in a Coulomb field. The constants
involved in the transformation matrix S are determined in such a way
that the resulting second - order radial differential equations do not
include fist - derivatives  of the 4-vector and/or of the scalar
potential. One can thus put the transformed radial equations in
suggestive forms for which their solutions can be inferred from the
known nonrelativistic solution of the Coulomb problem.

In Sec II the generalized Dirac - Coulomb equation is transformed
under a modified similarity transformation. Comparing the transformed
radial equations with the Schr$\ddot{o}$dinger - Coulomb problem we extract
the exact energy eigenvalues of the generalized Dirac - Coulomb
equation. In the same section we discuss various special cases
concerning the exact energy eigenvalues. We conclude in Sec III.

\section{A modified similarity transformation for the generalized
Dirac-Coulomb equation.}

The generalized Dirac-Coulomb equation involves a Coulomb potential in
the form of a superposition of the Lorentz - vector and Lorentz -
scalar terms $V_{v}(r)=-A_{1}/r$ and $V_{s}(r)=-A_{2}/r$, respectively. The
Dirac equation thus reads ( with the units $\hbar=c=1$)
\begin{equation}
H\Psi=E\Psi,
\end{equation}
where\\
\begin{equation}
H=\vec{\alpha}\cdot\vec{p}+\beta(m-A_{2}/r)-A_{1}/r,
\end{equation} \\
and the Dirac matrices $\vec{\alpha}$ and $\beta$ have their usual
meanings. Applying a similarity transformation [ 1, 2] to the Dirac
equation one gets\\
\begin{equation}
H^{'}\Psi^{'}=E\Psi^{'},
\end{equation}
with\\
\begin{equation}
H^{'}=SHS^{-1},
\end{equation} \\
\begin{equation}
\Psi^{'}=S\Psi,
\end{equation}
and\\
\begin{equation}
S=a+ib\beta\vec{\alpha}\cdot\hat{r},
\end{equation} \\
where $\hat{r}$ is the unit vector $\vec{r}/r$ and $a$ and $b$ are
real constants to be determined. For the central potentials above, the
transformed wave function is given by\\
\begin{equation}
\Psi^{'}= \left[ \begin{array}{c}
i R(r) \Phi^{l}_{jm}\\
Q(r) \vec{\sigma}\cdot\hat{r} \Phi^{l}_{jm}
\end{array} \right].
\end{equation}
In a straightforward manner one can calculate\\
\begin{equation}
E\Psi^{'}=SHS^{-1}\Psi^{'},
\end{equation}
to obtain two coupled equations for $R(r)$ ( the upper component) and
$Q(r)$ ( the lower component);\\
\begin{equation}
\left[ \begin{array}{ll}
M_{11} & M_{12}\\
M_{21} & M_{22}
\end{array} \right] \left[ \begin{array}{c}
R(r)\\
Q(r) \end{array} \right]=E \left[ \begin{array}{c}
R(r)\\
Q(r) \end{array} \right],
\end{equation}
with\\
\begin{equation}
M_{11}=(m-A_{2}/r)cosh \theta + (\partial_{r}+1/r)sinh \theta - A_{1}/r,
\end{equation}
\begin{equation}
M_{12}=-[(m-A_{2}/r)sinh \theta + (\partial_{r}+1/r)cosh \theta -
K/r],
\end{equation}
\begin{equation}
M_{21}=(m-A_{2}/r)sinh \theta + (\partial_{r}+1/r)cosh \theta +
K/r,
\end{equation}
\begin{equation}
M_{22}=-[(m-A_{2}/r)cosh \theta + (\partial_{r}+1/r)sinh \theta +
A_{1}/r],
\end{equation}
where $K=\tilde{\omega}(j+1/2)$, $\tilde{\omega}=\mp1$ for $l=j\mp1/2$,
 $cosh \theta=(a^{2}+b^{2})/(a^{2}-b^{2})$, and
$sinh \theta=2ab/(a^{2}-b^{2})$. The coupled equations for $R(r)$ and
$Q(r)$ are \\
\begin{equation}
M_{11}R(r)+M_{12}Q(r)=ER(r),
\end{equation}
and
\begin{equation}
M_{21}R(r)+M_{22}Q(r)=EQ(r).
\end{equation}
Multiply Eq(14) by $sinh \theta$, Eq(15) by $cosh \theta$ and subtract
to obtain\\
\begin{equation}
Q(r)=\frac{1}{\xi_{1}}[\partial_{r} + \frac{1}{r}+\frac{K}{r}cosh \theta
+ \frac{A_{1}}{r}sinh \theta + Esinh \theta]R(r),
\end{equation} \\
Next, multiply Eq(14) by $cosh \theta$, Eq(15) by $sinh \theta$ and
subtract to get\\
\begin{equation}
R(r)=\frac{1}{\xi_{2}}[\partial_{r} + \frac{1}{r}-\frac{K}{r}cosh \theta
- \frac{A_{1}}{r}sinh \theta - Esinh \theta]Q(r),
\end{equation}
where\\
\begin{equation}
\xi_{1}=m-\frac{A_{2}}{r}+\frac{A_{1}}{r}cosh \theta +
\frac{K}{r}sinh \theta + Ecosh \theta,
\end{equation}
and
\begin{equation}
\xi_{2}=m-\frac{A_{2}}{r}-\frac{A_{1}}{r}cosh \theta -
\frac{K}{r}sinh \theta - Ecosh \theta.
\end{equation}
Incorporating the regular asymptotic behaviour of the radial functions
near the origin; i.e. $R(r)\sim a_{1}r^{\gamma-1}$ and
$Q(r)\sim a_{2}r^{\gamma-1}$ at $r\sim0$, where the constant terms
proportional to mass and energy can be neglected, one gets\\
\begin{equation}
\gamma=\sqrt{K^{2}-A_{1}^{2}+A_{2}^{2}}.
\end{equation}
The positive sign of the square root has been chosen to allow not only
the normalization of the wave functions, but also the expectation
value of each partial operator within the transformed Hamiltonian.

To attain a great simplification in solving the radial equations,
Eq(16) and (17), one may choose\\
\begin{equation}
sinh \theta=[|K|A_{2}-\tilde{\omega}A_{1}\gamma]/[K^{2}-A_{1}^{2}],
\end{equation}
and
\begin{equation}
cosh \theta=[|K|\gamma-\tilde{\omega}A_{1}A_{2}]/[K^{2}-A_{1}^{2}].
\end{equation}
In fact one gets\\
\begin{equation}
\xi_{1}=m-A_{2}/r+\tilde{\omega}A_{2}/r+Ecosh \theta,
\end{equation}
and\\
\begin{equation}
\xi_{2}=m-A_{2}/r-\tilde{\omega}A_{2}/r-Ecosh \theta.
\end{equation}
Provided that\\
\begin{equation}
\tilde{\omega}= \left[ \begin{array}{llll}
+1 & for & K>0; & n=n_{r}+|K|+1\\
-1 & for & K<0; & n=n_{r}+|K| \end{array} \right.
\end{equation}
where $n_{r}$ and $n$ are the radial and principle quantum numbers,
respectively.

Moreover, for $\tilde{\omega}=+1$, Eq(23) and (24) read\\
\begin{equation}
\xi_{1}=m+Ecosh \theta,
\end{equation}
and\\
\begin{equation}
\xi_{2}=m-2A_{2}/r+Ecosh \theta,
\end{equation}
which when substituted in (16) and (17), eliminating $Q(r)$, imply\\
\begin{equation}
[E^{2}-m^{2}]R(r)=\left[-\partial_{r}^{2} - \frac{2}{r}\partial_{r} +
\frac{(\gamma^{2}+\gamma)}{r^{2}} - \frac{2(mA_{2}+A_{1}E)}{r} \right]
R(r).
\end{equation}

For $\tilde{\omega}=-1$, Eq(23) and (24) read\\
\begin{equation}
\xi_{1}=m-2A_{2}/r+Ecosh \theta,
\end{equation}
and\\
\begin{equation}
\xi_{2}=m-Ecosh \theta,
\end{equation}
which in turn when substituted in (16) and (17), eliminating $R(r)$,
imply\\
\begin{equation}
[E^{2}-m^{2}]Q(r)=\left[-\partial_{r}^{2} - \frac{2}{r}\partial_{r} +
\frac{(\gamma^{2}+\gamma)}{r^{2}} - \frac{2(mA_{2}+A_{1}E)}{r} \right]
Q(r).
\end{equation}

The first derivatives can be removed by defining $R(r)=r^{-1}\phi(r)$
and $Q(r)=r^{-1}q(r)$ to obtain\\
\begin{equation}
[E^{2}-m^{2}]\phi(r)=\left[-\partial_{r}^{2} +
\frac{(\gamma^{2}+\gamma)}{r^{2}} - \frac{2(mA_{2}+A_{1}E)}{r} \right]
\phi(r)
\end{equation}
for $\tilde{\omega}=+1$, and\\
\begin{equation}
[E^{2}-m^{2}]q(r)=\left[- \partial_{r}^{2} +
\frac{(\gamma^{2}+\gamma)}{r^{2}} - \frac{2(mA_{2}+A_{1}E)}{r} \right]
q(r)
\end{equation}
for $\tilde{\omega}=-1$.

It is obvious that Eqs(32) and (33) are in a form nearly identical to
that of the corresponding radial wave equation of Schr$\ddot{o}$dinger in the
Coulomb field [1-4, 8-11]. Their solutions can thus be inferred from
the known Schr$\ddot{o}$dinger - Coulomb solution [ 9-11]. Moreover, the
choices of $sinh \theta$ and $
cosh \theta$, consequently the S
transformation, have successfully transformed the coupled radial
equations of the generalized Dirac - Coulomb equation into simple
forms which are exactly soluble.

Inferred from the known nonrelativistic solution of the Coulomb problem
the exact energy eigenvalues for Eqs(32) and (33) are given through
the relation \\
\begin{equation}
E^{2}-m^{2}=-[mA_{2}+A_{1}E]^{2}/\tilde{n}^{2},
\end{equation}
which implies\\
\begin{equation}
\frac{E}{m}=-\frac{A_{1}A_{2}}{\tilde{n}^{2}+A_{1}^{2}} \pm \left[ \left(
\frac{A_{1}A_{2}}{\tilde{n}^{2}+A_{1}^{2}} \right)^{2} +
\frac{(\tilde{n}^{2}-A_{2}^{2})}{\tilde{n}^{2}+A_{1}^{2}}
\right]^{1/2}.
\end{equation}
Provided $\tilde{n}=n_{r}+\gamma+1$, where $n_{r}$ and $\gamma$ are
defined through Eqs(25) and (20), respectively. Eq(35) represents the
exact energy eigenvalues of the Dirac equation with an attractive
central potential that contains both the time component of a Lorentz
4-vector term,
$V_{v}(r)=-A_{1}/r$, and the Lorentz - scalar term,
$V_{s}(r)=-A_{2}/r$ [9]. It should be noted that in the case of the
superposition of the central potentials above $K^{2}$ is required to
be larger than $A_{1}^{2}-A_{2}^{2}$ otherwise $\gamma$ becomes
imaginary, causing the breakdown of the bound state solution.
Moreover, this result agrees with that obtained by Tutik [8] when only
the positive energy ( the particle case) solution is considered.

In the context of Eq(35) various special cases deserve attention.

For $A_{1}=0$ and $A_{2}\neq0$, $\gamma=\sqrt{K^{2}+A_{2}^{2}}$, and
\begin{equation}
E=\pm\left[1-\frac{A_{2}^{2}}{\tilde{n}^{2}} \right]^{1/2}.
\end{equation}
Which is the exact eigenvalue spectrum of the three - dimensional
Dirac equation with the scalar coupling potential $V_{s}(r)=-A_{2}/r$ [9].

For $A_{2}=0$ and $A_{1}\neq0$, $\gamma=\sqrt{K^{2}-A_{1}^{2}}$,\\
\begin{equation}
\tilde{n}=\frac{A_{1}E}{[m^{2}-E^{2}]^{1/2}}>0,
\end{equation}
and\\
\begin{equation}
E=m\left[1+\frac{A_{1}^{2}}{\tilde{n}^{2}}\right]^{-1/2}.
\end{equation}
Where the negative square root is not possible for it yields a
contradiction to Eq(37). Obviously this result is identical with the
known Sommerfeld's fine - structure formula [8,9].

For $A_{1}=A_{2}=A$, $\gamma=|K|$ and \\
\begin{equation}
E=m\left[\frac{-A^{2}}{\tilde{n}^{2}+A^{2}} \pm
\frac{\tilde{n}^{2}}{\tilde{n}^{2}+A^{2}}\right].
\end{equation}
The negative sign yields a solution $E=-m$ which is invalid for it
contradicts Eq(34). For the positive sign, it follows that \\
\begin{equation}
E=m\left[1-\frac{2A^{2}}{\tilde{n}^{2}+A^{2}}\right].
\end{equation}
This result allows positive and negative energy solutions.
However, Eq(34) suggests that positive and negative energy solutions
are possible if and only if $\tilde{n}^{2}>A^{2}$
 and $\tilde{n}^{2}<A^{2}$, respectively, otherwise impossible for
 they contradict Eq(34).
Moreover, it is clear from Eq(40) that at the limit
$A\rightarrow\infty$ the energy $E$ approaches $-m$ asymptotically
and contradicts Eq(34). 

\section{ Conclusions and remarks}

In this paper we have modified the similarity transformation S used by
Su [1] and Wong [2] to obtain exact energy eigenvalues of the
generalized Dirac - Coulomb equation. The modified transformation
reduces to that used by Su [1] and Wong [2] when the scalar potential
$V_{s}(r)$ is switched off.

To the best of our knowledge, this is the first time that a similarity
transformation is used to obtain the exact energy eigenvalues of the
generalized Dirac - Coulomb equation. The importance of this new
representation of the transformation S, as suggested by Eqs (28) and
(31), lies in the fact that it enables us to treat the generalized
Dirac - Coulomb radial functions $R(r)$ and $Q(r)$ as the precise
analogs to the radial wave functions of the nonrelativistic Coulomb
problem. Provided, as evident from Eqs(28) and (31), that the integer
orbital angular momentum of the radial Schr$\ddot{o}$dinger - Coulomb
equation becomes in the relativistic Coulomb problem a noninteger
( irrational) orbital angular momentum; i.e.
$l_{Schr\ddot{o}dinger} \equiv \gamma_{Dirac-Coulomb}$.

\end{document}